# Data analytics accelerates the experimental discovery of new thermoelectric materials with extremely high figure of merit

*Yaqiong Zhong[1#], Xiaojuan Hu[2#], Debalaya Sarker[3#], Qingrui Xia[1,5], Liangliang Xu[4], Chao Yang[1,5], Zhong-Kang Han[3]\*, Sergey V. Levchenko[3]\* and Jiaolin Cui[1]\**

[1] School of Materials and Chemical Engineering, Ningbo University of Technology, Ningbo 315211, China.
[2] Fritz-Haber-Institute of the Max Planck Society, Berlin 14195, Germany.
[3] Center for Energy Science and Technology, Skolkovo Institute of Science and Technology, Moscow 413026, Russia.
[4] Multidisciplinary Computational Laboratory, Department of Electrical and Biomedical Engineering, Hanyang University, Seoul 04763, Korea.
[5] School of Materials Science and Engineering, China University of Mining and Technology, Xuzhou 221116, China
[#] *These authors contributed equally.*

*Correspondence Email: S.Levchenko@skoltech.ru; H.Zhongkang@skoltech.ru; cuijl@nbut.edu.cn*

**Abstract**

Thermoelectric (TE) materials are among very few sustainable yet feasible energy solutions of present time. This huge promise of energy harvesting is contingent on identifying/designing materials having higher efficiency than presently available ones. However, due to the vastness of the chemical space of materials, only its small fraction was scanned experimentally and/or computationally so far. Employing a compressed-sensing based symbolic regression in an active-learning framework, we have not only identified a trend in materials' compositions for superior TE performance, but have also predicted and experimentally synthesized several extremely high performing novel TE materials. Among these, we found $Cu_{0.45}Ag_{0.55}GaTe_2$ to possess an experimental figure of merit as high as ~2.8 at 827 K, which is a breakthrough in the field. The presented methodology demonstrates the importance and tremendous potential of physically informed descriptors in material science, in particular for relatively small data sets typically available from experiments at well-controlled conditions.

The ever-increasing energy demand of present era and the resulting limitless combustion of fossil fuels have already lifted global pollutant levels to an alarming threshold. Scavenging waste heat with thermoelectric generators is one of the few viable ways towards sustainable energy revolution. In thermoelectric materials a temperature gradient produces an electric potential, giving us a handle to convert waste heat to electricity.[1-4] However, the efficiency of such a conversion remained too low for its economically viable large-scale utilization. The progress in the field was limited by the need of optimizing a variety of conflicting parameters. A large Seebeck coefficient ($\alpha$), a low thermal conductivity ($\kappa$), and a high electrical conductivity ($\sigma$) are required to maximize the thermoelectric figure of merit $zT = \alpha^2\sigma T/\kappa$. Here $\kappa$ is the sum of electronic ($\kappa_e$) and lattice ($\kappa_L$) components of thermal conductivity. The interdependence of the TE transport properties implies the crucial importance of synergistic optimization of the electronic and thermal transport properties to achieve high zT values.

Excellent thermoelectric performances have been achieved in different types of thermoelectric materials due to the devotement of a substantial amount of research efforts to identify material compositions and configurations with optimized transport characteristics providing best possible TE performance. Among them, half-Heusler alloys show great potential for high-temperature power generation applications owing to their high stability at high temperature (a high zT of ~1.5 at 1200 K was achieved in p-type NbFeSb).[5] Inorganic $Ba_8Ga_{16}Sn_{30}$ clathrates have achieved high peak zT of 1.45 at around 500 K which indicates their great potential for low- and medium-temperature applications.[6] Skutterudites achieved a high zT of ~1.8 at 823 K for (Sr, Ba, Yb)$_y$Co$_4$Sb$_{12}$ + 9.1 wt% In$_{0.4}$Co$_4$Sb$_{12}$.[7] In recent years, chalcogenides were identified as very promising thermoelectric materials due to the discovery of a few members of this class with extremely high thermoelectric performance (zT > 2).[8-16] In particular, the binary chalcogenides, including PbTe, $Cu_2Se$, GeTe, and SnSe, were "superstars" in the thermoelectric family.[9-14] The ternary chalcogenide compounds are also extensively investigated due to their large variety and high TE performance (zT ~1.9 at 585 K was achieved in $AgSb_{0.96}Zn_{0.04}Te_2$).[15, 16] The mechanisms governing the performance of TE materials were also proposed. While substitution of Te by Se in PbTe has reportedly enhanced the zT by increasing phonon scattering,[17] doping with thallium has improved the TE performance of PbTe by introducing new states in the band gap and thereby improving the Seebeck coefficient.[18] The "liquid-like" behavior of copper ions around the Se sublattice in $Cu_{2-x}Se$ results in low lattice thermal conductivity, which enables high zT.[19] The n-type SnSe single crystals have an extremely high zT of ~2.8 at 773 K by Br-doping via well-controlled synthesis techniques due to a unique 3D charge and 2D phonon transport behavior.[13] The polycrystalline phases of SnSe, which are more easily obtained experimentally, also possess a high zT value of ~1.8 at 816 K.[20] Despite all these substantial efforts over the years, only a handful of materials and their optimized working conditions (i.e. temperatures) for best performances have been identified hitherto. Thus, a quick yet reliable means to scan the huge compositional space of TE materials is essential for the accelerated discovery of new high-performance TE materials.

With the advance of computational resources, first-principles calculations have been proven very useful not only for the initial screening process but also for understanding and interpretation of experimental results.[21, 22] Electronic transport properties of more than 48,000 inorganic materials have been calculated and reported as a database to be used in many fields from thermoelectricity to electronics and photovoltaics.[23] While a lot of high-throughput screening with density-functional theory (DFT) inputs have been performed to identify materials with high power factors or low thermal conductivities, including transition-metal oxides, nitrides, sulphides, etc., the actual performance and stability of the predicted materials have remained unconfirmed.[24-26] This is unfortunately a typical situation in many fields of materials science: powerful theoretical and data analytics methodologies are intensively developed and employed to screen materials space, but experimental confirmation is very scarce. The reason is not only the relatively high cost of experiments, but also a lack of effort to produce consistent experimental datasets with well-documented metadata (including experimental conditions, materials synthesis protocol, etc.). Recently an attempt to address this problem systematically has been made in the field of heterogeneous catalysis.[27] There have been several attempts in developing experimental databases in the field of thermoelectricity,[28] but the sparsity of the available datasets make the screening either biased or wrong.

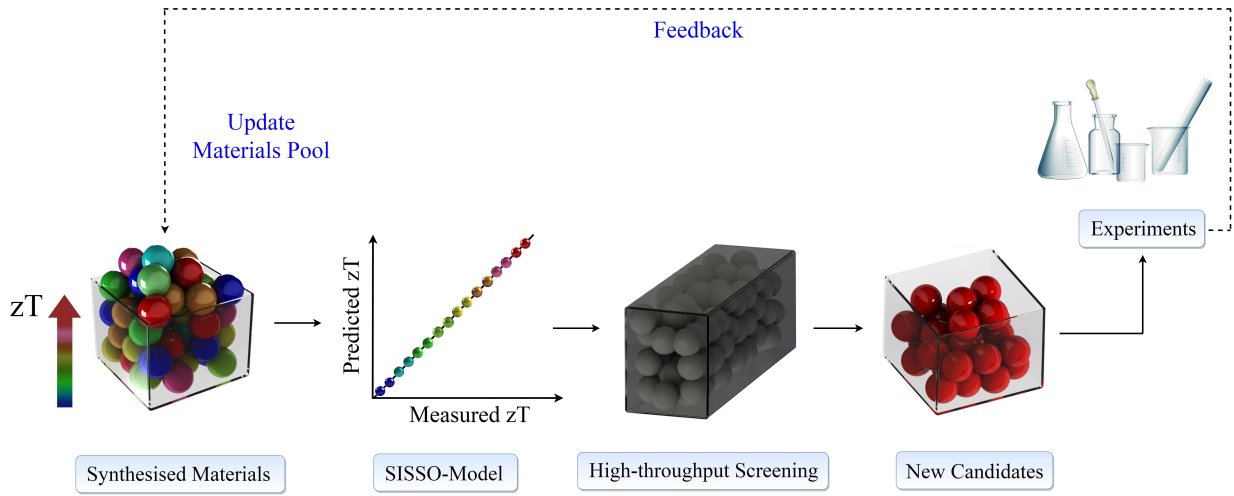

**Figure 1.** Schematic representation of materials design with active learning.

Herein, we generate ~600 data points for in-house experimentally synthesized ternary chalcogenide materials A-B-$C_2$ (where A = Cu or Ag; B = Bi, In, Sb, or Ga; C = Se or Te) with proper doping to ensure consistency of environmental and instrumental effects. Using the recently developed compressed-sensing based data analytics approach SISSO (sure independence screening and sparsifying operator),[29] we have identified descriptors that can be used to predict the TE performance of ternary A-B-$C_2$ type chalcogenide materials in a fast yet reliable manner. SISSO enables us to identify the best low-dimensional descriptors (sets of descriptive parameters of materials that are either known or can be easily obtained) in an immensity of offered candidates. The search for materials with optimized zT values is performed in an active learning framework. Starting with an initial pool of experimental data, we first identify a descriptor and predict new candidates. Followed by that, we synthesize few of the predicted high-performance TE materials and include them in the data pool. This procedure (Figure 1) has been repeated several times until the SISSO model has converged.

Thus, our study goes beyond the traditional expensive and time-consuming intuition-driven or trial-and-error experimental/theoretical approaches. It reveals a relationship between TE figure of merit (zT), elemental composition, and physical features of atoms of involved species. Using data driven active learning, we have successfully predicted and experimentally verified several new TE materials of the ternary A-B-$C_2$ type chalcogenide family, which are not only high-performing but also are stable over a broad temperature range.

**Results**

Around 600 data points (zT values) from experimentally synthesized ternary A-B-$C_2$ type chalcogenide compounds within the temperature window of 300-850K are used as the training dataset. The in-house synthesis provides our dataset the consistency necessary for machine learning (ML). In the SISSO method, a huge pool of candidate features of increasing complexity is first constructed iteratively by applying a set of mathematical operators to a set of primary features. The primary features include only temperature and experimentally known materials properties listed in Table 1. The target property (zT) is expressed as a linear combination of the complex features $d_i$:

$$zT = \sum_{i=1}^{N} c_i d_i + A,$$

The set of $d_i$ is called a descriptor with dimension $N$. $A$ is the intercept. Compressed sensing is used to identify best model and the corresponding descriptor for each dimension up to a maximum value. The initial choice of primary features is crucial for the predictive performance of descriptors identified by SISSO. Materials properties such as atomic species and their relative concentrations,[30, 31] atomic radii,[32, 33] atomic weights,[33] electronegativities,[34, 35] ionization energies,[36] and heats of fusion/vaporization[37-39] are reported to have strong impact on TE performance of different thermoelectric materials. Therefore, we have constructed the primary feature space with these properties.

**Table 1.** Primary features used for the descriptor construction for the in-house experimentally synthesized ternary based chalcogenide materials A-B-C$_2$ (where A = Cu or Ag; B = Bi, In, Sb, or Ga; C = Se or Te) with proper doping.

| Primary Feature | Element | Symbol |
|---|---|---|
| Temperature (K) | – | T |
| Dopant concentration | | $C_{A*}, C_{B*}, C_{C*}$ |
| Electronegativity (eV) | | $EN_A, EN_{A*}, EN_B, EN_{B*}, EN_C, EN_{C*}$ |
| Ionization Energy (eV) | | $IE_A, IE_{A*}, IE_B, IE_{B*}, IE_C, IE_{C*}$ |
| Heat of fusion (eV) | A, A*, B, B*, C, C* | $HF_A, HF_{A*}, HF_B, HF_{B*}, HF_C, HF_{C*}$ |
| Heat of vaporization (eV) | | $HV_A, HV_{A*}, HV_B, HV_{B*}, HV_C, HV_{C*}$ |
| Atomic Radius (Å) | | $AR_A, AR_{A*}, AR_B, AR_{B*}, AR_C, AR_{C*}$ |
| Atomic Weight (a.u.) | | $AW_A, AW_{A*}, AW_B, AW_{B*}, AW_C, AW_{C*}$ |

The dopants at A, B, and C sites are represented by A* (A* = Cu, Ag, Zn, or Na), B* (B* = Ga, In, Bi, Sb, Zn, or Sn), and C* (C* = Te or Cl), respectively. $C_{M*}$ (M = A, B, C) are concentrations of dopants defined as the fraction of corresponding sites occupied by M*. Minority species ($C_{M*} \leq 0.5$) are always considered as dopant, so that $C_{M*}$ varies between 0 and 0.5. When the concentration of a dopant is zero ($C_{M*} = 0$), the values of $EN_{M*}$, $IE_{M*}$, $HF_{M*}$, $HV_{M*}$, $AR_{M*}$, and $AW_{M*}$ are equal to the values of $EN_M$, $IE_M$, $HF_M$, $HV_M$, $AR_M$, and $AW_M$.

In SISSO overfitting may occur with increasing dimensionality of the descriptor. To avoid overfitting, 10-fold cross-validation (CV10)[40] is employed to identify the optimal dimension of the model. For each cycle displayed in Figure 1, the materials pool is first split into 10 subsets (the materials datasets for each cycle are collected in the Supplementary Data1), and the descriptor identification along with the model training is performed using 9 subsets. Then the error in predicting properties of the systems in the remaining subset is evaluated with the obtained model. The CV10 error is calculated as the average value of the test errors obtained for these ten subsets. The results for each cycle are collected in Figure S1. As can be seen in Figure S1, for each cycle the CV10 error reduces gradually from $N = 1$ to $N = 7$ (1D to 7D) descriptors, which suggests that overfitting does not occur for dimensions up to 7D. This is due to the relatively large number of training data points for SISSO (although it is still small compared to typical dataset sizes needed for traditional ML approaches such as neural networks). The root-mean-square fitting error (RMSE) for the 7D descriptor for each cycle is already small and is only slightly smaller than for the 6D descriptor. Since higher dimensions mean higher model complexity and computational cost of training, we have carried out all our analysis with up to 7D descriptors. To further confirm the predictive power of our best model (7D) at the final iteration, we have validated the model based on the zT values from previous reports on chalcogenide materials from other groups in Table S1 alongside with the SISSO-predicted values. Despite experimental uncertainties, the predictions remain overall consistent with experiment.

The overall good performance of our best model at the final iteration is demonstrated by low RMSE in zT (0.14). The distribution of errors in predicted zT for different temperature ranges, different zT ranges, and the overall distribution are shown in Figures 2Sa-c. Figure 2 shows the distribution of zT for different temperature ranges in the final dataset. The top five largest deviations between SISSO model predicted zT values and the experimentally measured zT values are collected in Table S2. The maximum absolute error reaches 0.76 for $Ag_{0.55}Cu_{0.45}GaTe_2$ at 730.3 K. In general, the absolute errors larger than 0.6 are all from $Cu_{1-x}Ag_xGaTe_2$ systems with experimentally measured zT values larger than two at temperature larger than 730 K. Fewer data points measured at high temperatures (>750K) and high-zT materials (> 1.2) explain why the prediction errors are larger for these cases. However, for all the materials the SISSO model underestimates zT, i.e., all materials predicted to have good thermoelectric properties can be expected to have even better properties in reality.

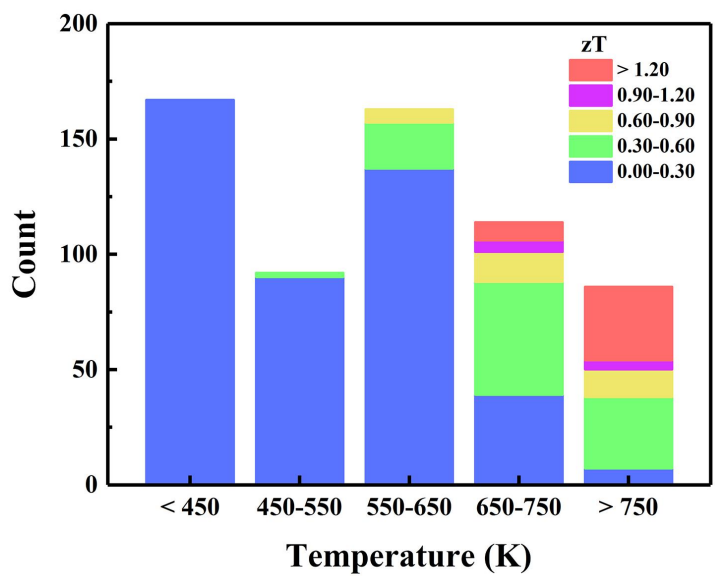

**Figure 2.** Distribution of the data within the final dataset for different temperature ranges and zT ranges.

The descriptor components $d_i$, the coefficients $c_i$, importance score, and the occurrence number of each component during the CV10 processes for the best SISSO model at the final iteration are given in Table 2 (the results for other iterations are collected in Table S3-5). One of the advantages of SISSO over other data analytics approaches is the physical interpretability of the models. By inspecting the descriptor components, one can see how SISSO selects physically meaningful correlations between the target property (zT) and the combinations of primary features. In particular, occurrence of $T^2$ in $d_1$ reflects the fact that $zT/T = \alpha^2\sigma/\kappa$ consistently increases with temperature for the materials in the considered class. The occurrence of $1-C_{A*}$ in the denominator of $d_1$ indicates that TE efficiency increases with increasing disorder in the A-site, which reduces thermal conductivity.[4, 41, 42] It should be noted that the dependence of zT on the dopant concentration is strongly nonlinear. A small dopant concentration can have a strong effect on the TE performance due to the formation of impurity band within the gap.[18] The dependence of $d_1$ and $d_2$ on atomic features indicates that zT is increased for smaller and lighter atoms at the B-site, with at least the minority species (B*) being as light as possible relative to the majority species at the C-site. According to $d_4$, additional gain in zT can be achieved by choosing B atoms lighter than majority of atoms at A sites. The disparity of size and weight between B and C/A atoms maximizes the rattling effect, which is responsible for a decrease of thermal conductivity in clathrates.[43]

At the same time, smaller atomic radius of majority species at site B and disorder at site A are detrimental for electrical conductivity, consistent with the dependence of $d_3$ on these features. This reflects the well-known dilemma in solid solutions that impurity scattering will reduce both the thermal conductivity and carrier mobility.[33, 44] We note that concentration of A* will only improve zT up to a limit, namely while the phonon scattering rises due to the lattice disorder but the carrier mobility remains unaffected.[17] In $Cu_{1-x-\delta}Ag_xInTe_2$, zT reportedly increases only until $x = 0.2$,[45] while in $Cu_{1-x}Ag_xGa_{0.6}In_{0.4}Te_2$ zT reaches its highest value of 1.64 (at 873K) for $x = 0.3$.[15] Thus, our data analysis shows that selecting impurity atoms with small radius is indeed an effective strategy to suppress lattice thermal conductivity and at the same time maintain the carrier mobility.

The appearance of the heat of vaporization $HV_A$ in $d_3$ can be related to the ability of element A with higher heat of vaporization to form stronger bonds with surrounding atoms and thus also reduce disorder and maintain higher electrical conductivity at elevated temperatures.[37-39] The appearance of electronegativity (EN) in $d_5$ accounts for the effects of electrical conductivity ($\sigma$) alongside with $\kappa$. The larger the $EN_B$ is compared to $EN_C$ the higher is the zT. This is due to the increase of the carrier mobility and carrier concentration induced by the ejection of electrons from element C. This is also consistent with previous findings that high electronegativity mismatch is beneficial for performance of thermoelectric alloys.[34]

**Table 2.** The descriptor components $d_i$, the coefficients $c_i$, importance score$^\$$, and the occurrence number of each component during the CV10 processes for the best SISSO model at the final iteration.

| | descriptor component | coefficient | importance score | occurrence number |
|---|---|---|---|---|
| $d_1$ | $T^2/((1-C_{A*}) \cdot AW_B)$ | 0.78685E-04 | 0.65 (0.61) | 10 |
| $d_2$ | $(T/AR_{B*}) \cdot \lvert AW_C - AW_{B*} \rvert$ | 0.13073E-04 | 0.38 (0.17) | 7 |
| $d_3$ | $((1-C_{A*}) \cdot AR_B \cdot HV_A)/T$ | 0.12837E+03 | 0.21 (0.14) | 1 |
| $d_4$ | $\lvert AW_B - AW_C \rvert / \lvert AW_A - AW_{B*} \rvert$ | -0.47799E-01 | 0.18 (0.16) | 1 |
| $d_5$ | $((1-C_{A*}) \cdot AR_{A*})/(EN_B - EN_C)$ | -0.74883E-01 | 0.04 (0.06) | 0 |
| $d_6$ | $(AW_A/T)/\lvert HV_B - HV_{A*} \rvert$ | 0.30740E-01 | 0.06 (0.03) | 0 |
| $d_7$ | $(T \cdot C_{B*})/(AR_A - AR_C)$ | -0.85704E-04 | 0.02 (0.01) | 2 |

$^\$$The importance score based on RMSE (MaxAE) for each descriptor component is calculated as follows. The component is removed from the descriptor, the model is refit with the remaining components, and RMSE and MaxAE are calculated. The score is calculated as 1 – RMSE (MaxAE) (all components) / RMSE (MaxAE) (all – 1 component). Score 0 means that removing the component does not change RMSE (MaxAE) of the model.

Since the model clearly captures the physics governing the performance of the thermoelectric materials, we proceed with the high-throughput (HT) screening of the huge compositional space of ternary based chalcogenide materials A-B-C$_2$ (see Supplementary Methods for details). Results of HT screening of more than 10,000 data points (zT values) from ternary A-B-C$_2$ type chalcogenide compounds are presented in Figure 3a. We have found many new promising TE materials that have not been realized experimentally before. Moreover, based on these results, we have synthesized the compound Ag$_{0.55}$Cu$_{0.45}$GaTe$_2$, showing an experimental zT value as high as ~2.8 at 827K, which have been repeated for several times (Figure S3). The results of thermal diffusivity are also reproduced from a third-party laboratory (see Figure S4 and Supplementary Appendix for details). The maximum zT values for other types of thermoelectric materials known to date are displayed in Figures 3b and c for comparison.

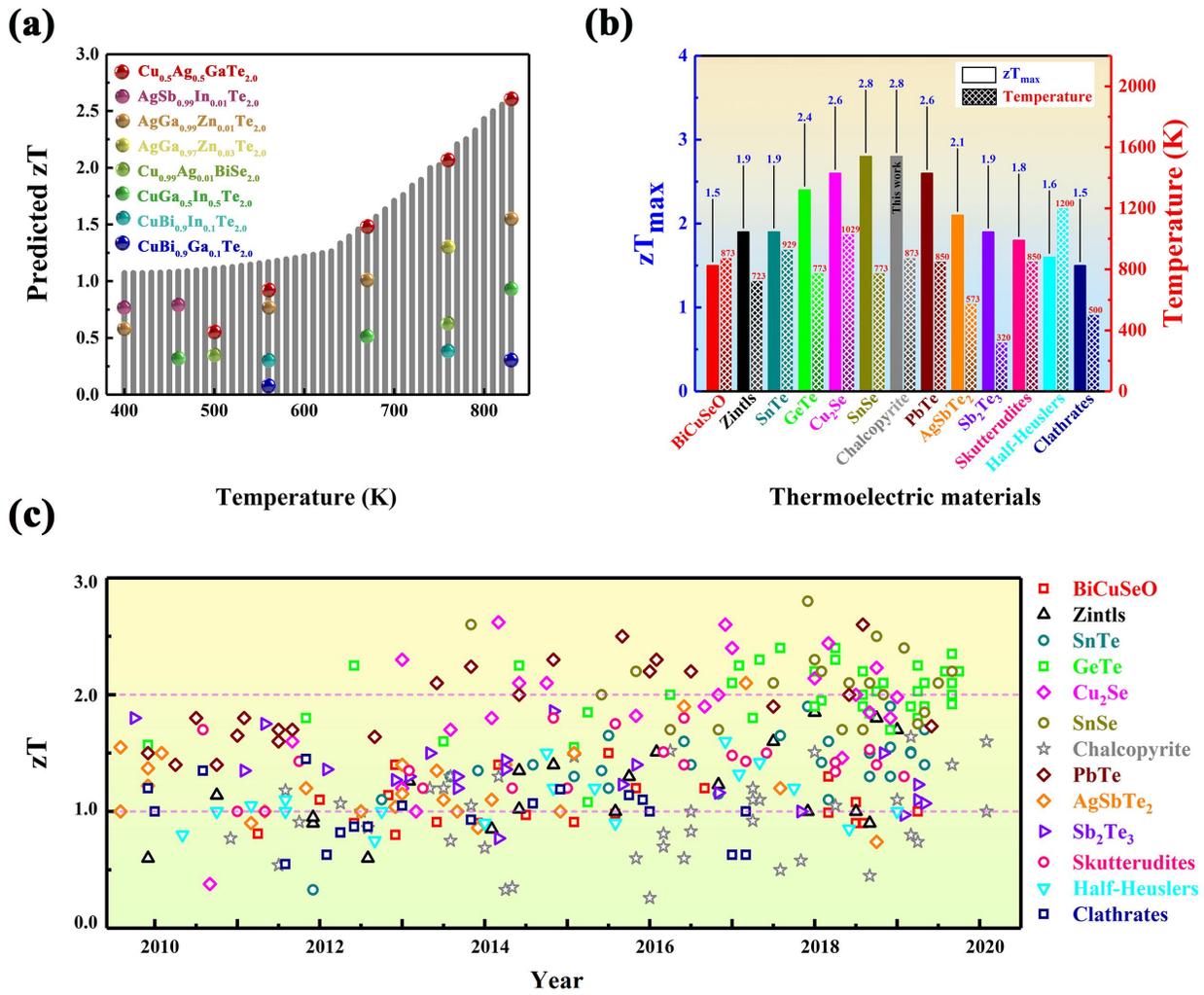

**Figure 3.** (a) Results of high-throughput search for new thermoelectric materials using the SISSO model. The height of the gray bar represents the maximum predicted zT value from the highthroughput screening. (b) Comparison of the experimental maximum zT values in other types of thermoelectric materials. (c) Experimental zT values of established systems since 2009 (the zT values and original literatures are collected in Supplementary Data2).

The phase stability of $Ag_{0.55}Cu_{0.45}GaTe_2$ can be seen from the XRD pattern in Figure S5. One can also see in Figure 4 that, although increasing the Ag concentration beyond 0.55 increases Seebeck coefficient, it does not reduce the lattice thermal conductivity any further. Rather, the additional Ag lowers the electrical conductivity so that zT decreases. It is exactly this interplay of various parameters that makes the search for high-performance thermoelectric materials so non-trivial.

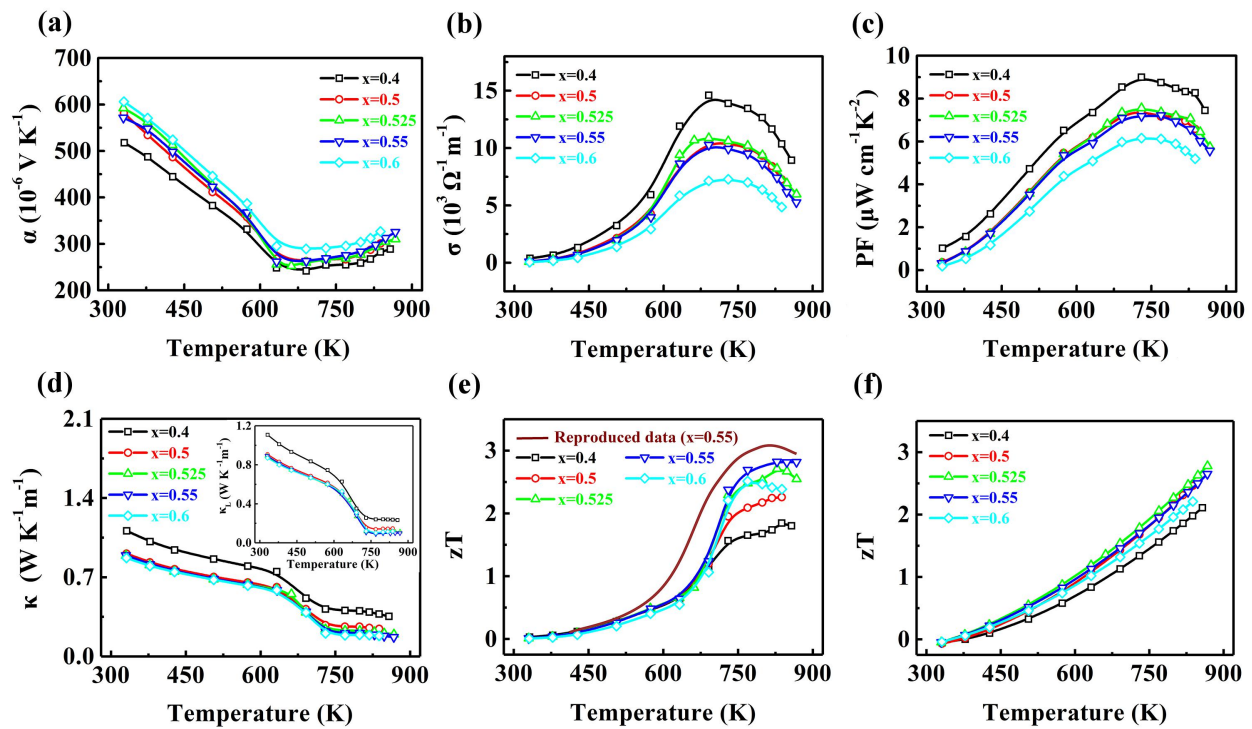

**Figure 4.** Measured parameters of the bulk $Cu_{1-x}Ag_xGaTe_2$ ($x$ = 0.4-0.6) as a function of temperature: (a) Seebeck coefficients ($\alpha$); (b) Electrical conductivities ($\sigma$); (c) Power factors (*PF*); (d) Total thermal conductivities ($\kappa$) and lattice thermal conductivities ($\kappa_L$); (e) Experimentally measured figure of merit (zT); (f) SISSO-predicted figure of merit (zT).

In summary, we have not only overcome one of the major hurdles of traditional TE material synthesis, namely exploring the very vast configurational space, but have also demonstrated the power of data analytics and machine learning in accelerating thermoelectric materials design. By learning from our own experimental dataset, obtained at consistent well-controlled conditions, we have developed a ML model that allows us to predict many new materials with exceptional TE performance effortlessly and reliably. Due to physical interpretability of our model, we were able to clearly disentangle interconnected, often conflicting effects of basic materials parameters on TE performance, and find a way to overcome the optimization difficulties by fully utilizing the large variability of the parameters within the explored materials class. The complex relationship between the primary features and the target property zT emphasizes the absolute necessity of advanced data analytics tools for the advancement of new functional materials discovery. Moreover, the successful synthesis and characterization of our predicted materials have paved out new lanes in TE research. In addition to the fact that these materials have exceptionally high zT values and stabilities and are therefore very promising for several TE applications, the success of our active-learning ML strategy clearly shows that such concerted approaches have a great potential for future materials design.

## Methods

### Data-analytics:

The descriptors are obtained with SISSO, using ~600 experimental zT inputs as training data. SISSO is meant to single out a simple yet physically intuitive descriptor from an immensely large set of candidates. Initially, a huge pool consisting of more than ten billion candidate descriptors, is constructed. Then an iterative approach is employed to search the descriptors by combining pre-defined primary features and a set of mathematical operators (+, −, ·, /, log, exp, exp−, $^{-1}$, $^{2}$, $^{3}$, $\sqrt{}$, $\sqrt[3]{}$, |−|). The complexity of the obtained descriptors depends on how many times the operators are employed. In the present study, we have considered feature space of complexity level up to two, $\Phi_1$ and $\Phi_2$.[29] Any given feature space of complexity $n$ ($\Phi_n$) also contains all of the lower (i.e. $n$-1) feature spaces. The details of the SISSO model identification procedures and the high-throughput screening of new materials/compositions at each iteration are given in the Supplementary Methods.

### Experimental:

### Sample preparation

Bulk samples of polycrystalline $Cu_{1-x}Ag_xGaTe_2$ ($x$=0-0.6) were prepared by vacuum melting–annealing combined with spark plasma sintering (SPS) using elemental Cu, Ag, Ga and Te (99.999%, Emei Semicon. Mater. Co., Ltd. Sichuan, CN). All the raw materials were weighed and mixed according to the above formulae and sealed in quartz tubes under vacuum, which was gradually heated up to 1273 K at a heating rate of 100 Kh$^{-1}$, and incubated for 28 h. Afterwards, the ampoules were slowly cooled to 873 K at a rate of 15 Kh$^{-1}$ followed by quenching in water and then dwelt at 813 K for 72 h. The obtained chunks were ball-milled into fine powders for 10 h, and then sintered by the spark plasma sintering apparatus (SPS-1030) at 673 K with a pressure of 55 MPa.

### Transport property measurements

The densified bulk samples of size ~2.5×3×12 mm$^3$ and $\phi$10×1.5 mm were prepared for electrical property and thermal diffusivity measurement. The Seebeck coefficients and electrical conductivities were performed with a ZEM-3 device (ULVAC-RIKO, Japan) under a helium atmosphere from room temperature to ~870 K with an uncertainty of < 5.0%. Thermal conductivity ($\kappa$) was calculated via $\kappa = DC_p\rho$, where the thermal diffusivity (D) was measured by the laser flash method (TC-1200RH, ULVAC-RIKO, Japan) with a precision of ~10.0% and confirmed by NETZSCH LFA457, Germany. The heat capacities ($C_p$) were estimated following the Dulong-Petit rule, $C_p = 3n$R (here n is the number of atoms per formula unit and R gas constant). The sample density ($\rho$) was measured by the Archimedes method. When calculating the electronic thermal conductivities ($\kappa_e$) according to the equation $\kappa_e = L\sigma T$, the Lorenz numbers L were estimated by using the formula $L=1.5+\exp(-|\alpha|/116)$ (where L is in $10^{-8}$ W$\Omega$K$^{-2}$ and $\alpha$ in $\mu$VK$^{-1}$).[46] The three physical parameters ($\alpha$, $\sigma$, and $\kappa$) were finalized after three measurements. The total uncertainty for zT is ~20%.

## Acknowledgments

The data-analytics methodology development is supported by RSF grant 21-13-00419. J.C. is supported by the National Natural Science Foundation of China (51671109).

## Author contributions

J.C. and Z.-K.H. created the idea and conceived the work. J.C., Z.-K.H., and S.V.L. designed and supervised the project. J.C. supervised the experimental synthesis and analysis. Y.Z. and Q.X. synthesized the materials and measured the properties of the thermoelectrics. X.H. and D.S. performed the data analytics. Y.Z., X.H., D.S., Z.-K.H., J.C., and S.V.L. co-wrote the manuscript. All authors contributed to the analysis and interpretation of the results. All the authors commented on the manuscript and have given approval to the final version of the manuscript.

## Competing interests

The authors declare no competing financial interests.

## Additional information

Supplementary Information is available for this paper at http://www.nature.com/nature.
Correspondence and requests for materials should be addressed to J.C. or Z.-K.H. or S.V.L.

# Data analytics accelerates the experimental discovery of new thermoelectric materials with extremely high figure of merit

Yaqiong Zhong[1#], Xiaojuan Hu[2#], Debalaya Sarker[3#], Qingrui Xia[1,5], Liangliang Xu[4], Chao Yang[1,5], Zhong-Kang Han[3]*, Sergey V. Levchenko[3]* and Jiaolin Cui[1]*

[1] School of Materials and Chemical Engineering, Ningbo University of Technology, Ningbo 315211, China.
[2] Fritz-Haber-Institute of the Max Planck Society, Berlin 14195, Germany.
[3] Center for Energy Science and Technology, Skolkovo Institute of Science and Technology, Moscow 413026, Russia.
[4] Multidisciplinary Computational Laboratory, Department of Electrical and Biomedical Engineering, Hanyang University, Seoul 04763, Korea.
[5] School of Materials Science and Engineering, China University of Mining and Technology, Xuzhou 221116, China
[#] These authors contributed equally.

Correspondence Email: S.Levchenko@skoltech.ru; H.Zhongkang@skoltech.ru; cuijl@nbut.edu.cn

## Supplementary Methods

For constructing the $\Phi_1$ and $\Phi_2$ feature spaces we made use of the set of algebraic/functional operators given in eq. 1.

$$\hat{H}^{(m)} \equiv \{+, -, \cdot, /, \log, \exp, \exp-, ^{-1}, ^2, ^3, \sqrt{}, \sqrt[3]{}, |-|\}, \qquad (1)$$

The superscript m indicates that when applying $\hat{H}^{(m)}$ to primary features $\varphi_1$ and $\varphi_2$ a dimensional analysis is performed, which ensures that only physically meaningful combinations are retained (e.g. only primary features with the same unit are added or subtracted). All primary features included in this study were obtained from the literature.[1] The values of the primary features for the training data sets at each iteration can be found in the file "Supplementary Data".

The sparsifying $\ell_0$ constraint is applied to a smaller feature subspace selected by a screening procedure (sure independence screening (SIS)), where the size of the subspace is equal to a user-defined SIS value times the dimension of the descriptor. The SIS value is not an ordinary hyperparameter and its optimization through a validation data set is not straightforward. Ideally, one would want to search the entire feature space for the optimal descriptor. However, this is not computationally tractable since the computational cost of the sparsifying $\ell_0$ constraint grows exponentially with the size of the searched feature space. Instead, the SIS value should be chosen as large as computationally possible. The reasonable SIS values were chosen based on the convergence of the training error.

The high-throughput (HT) screening of the huge compositional space of ternary based chalcogenide materials $A_{1-x}A^*_xB_{1-y}B^*_yC_{2-z}C^*_z$ (where A = Cu or Ag; B = Bi, In, Sb, or Ga; C = Se or Te; A* = Cu, Ag, Zn, or Na; B* = Ga, In, Bi, Sb, Zn, or Sn; C* = Te or Cl) was performed at the final iteration. The values of x, y, and z were limited to smaller than 0.1, 0.1, and 0.2, respectively, which corresponding to 10% of dopant concentration. The mixtures of $CuInTe_2$, $CuGaTe_2$, $AgInTe_2$, and $AgGaTe_2$ are stable in a large temperature range according to the literature.[2] Thus, their mixtures are also included in our high-throughput screening. The best ternary, quaternary, and quintenary A-B-C$_2$ type based chalcogenide materials were systematically screened at the first, second, and third iteration, respectively. The newly found best systems at each iteration are collected in the Supplementary Data.

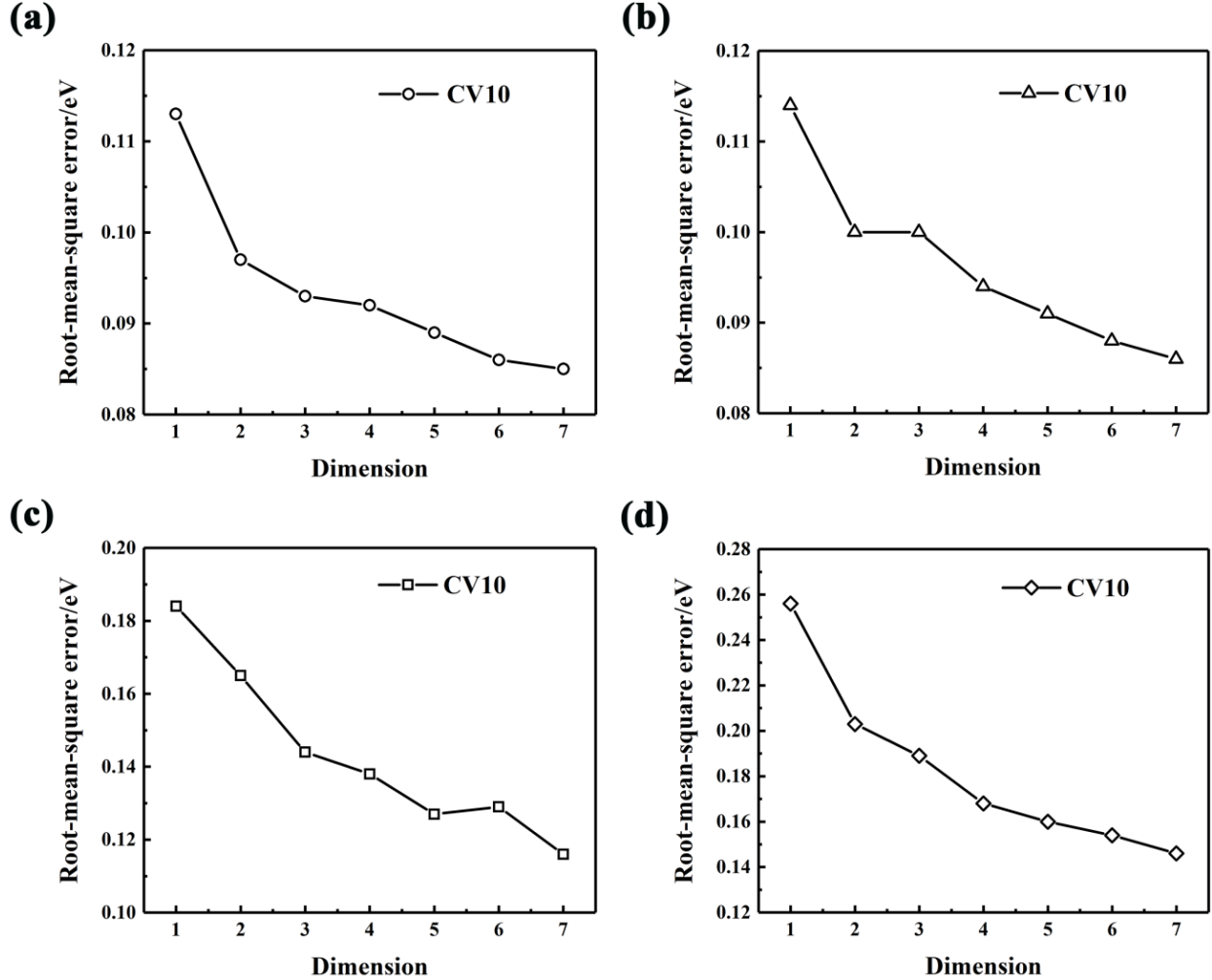

**Figure S1.** CV10 error at each iteration.

**Table S1.** zT values from previous reports on chalcogenide materials from other groups alongside with the SISSO-predicted values.

| systems | temperature (K) | SISSO-predicted | previous reported |
| --- | --- | --- | --- |
| CuGaTe$_2$ | 950 | 1.48 | 1.40[3] |
| Cu$_{0.7}$Ag$_{0.3}$In$_{0.6}$Ga$_{0.4}$Te$_2$ | 873 | 1.30 | 1.64[2] |
| AgSb$_{0.98}$Bi$_{0.02}$Se$_2$ | 680 | 0.66 | 1.15[4] |
| AgSbTe$_2$ | 533 | 0.81 | 1.55[5] |
| AgBiSe$_2$ | 700 | 0.46 | 1.50[6] |
| AgSb$_{0.96}$Zn$_{0.04}$Te$_2$ | 585 | 1.16 | 1.90[7] |
| AgSbTe$_{1.98}$Se$_{0.02}$ | 565 | 0.83 | 1.37[8] |
| AgSbTe$_{1.85}$Se$_{0.15}$ | 575 | 0.84 | 2.10[9] |
| AgSb$_{0.93}$In$_{0.07}$Te$_2$ | 650 | 0.88 | 1.35[10] |

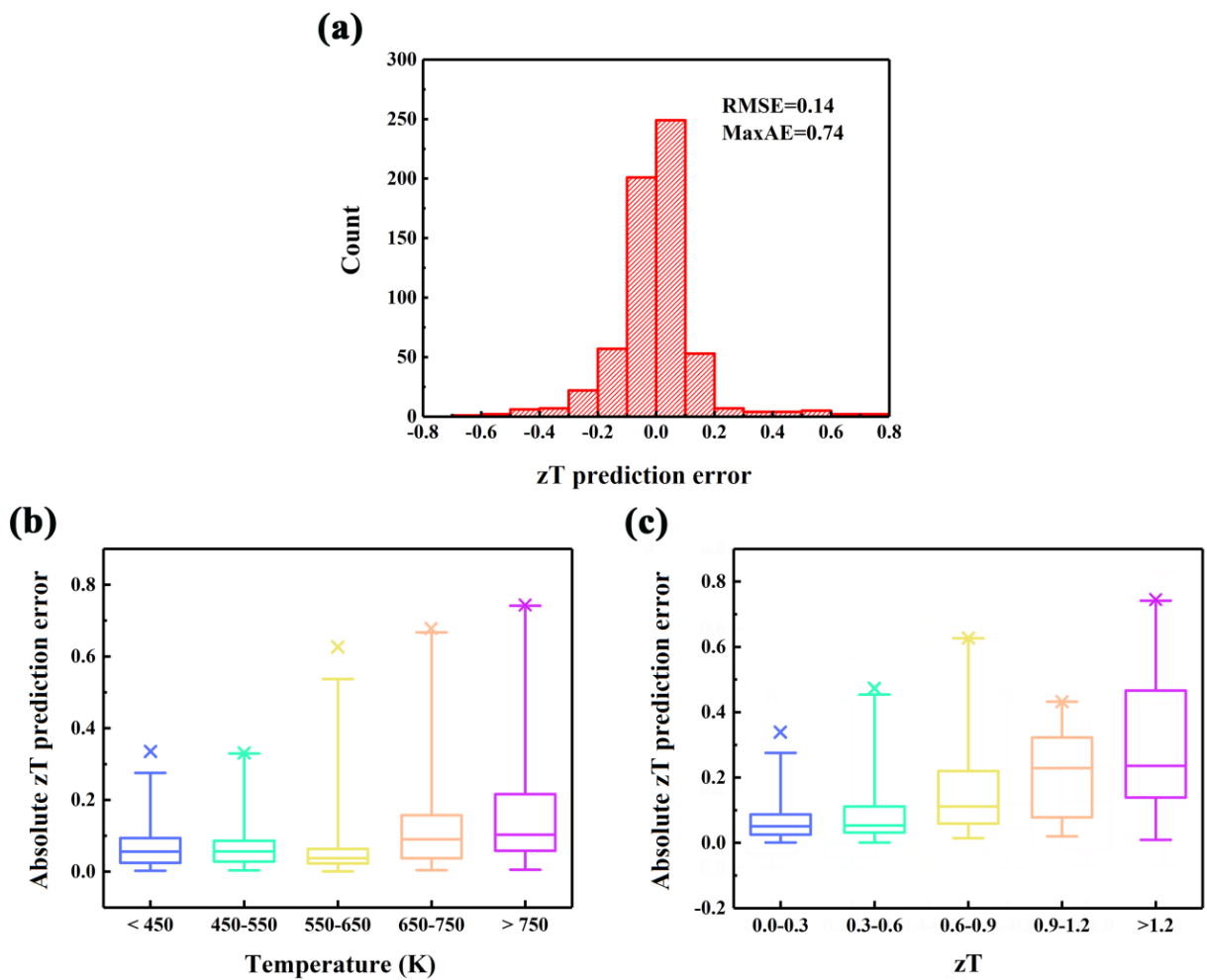

**Figure S2.** (a) Distribution of zT prediction errors for the best SISSO model at the final iteration; (b) and (c) Box plots for error distribution as a function of temperature range and zT range, respectively. The upper and lower limits of the rectangles show the 75th and 25th percentiles of the distributions, the internal horizontal lines mark the median (50th percentile), and the upper and lower limits of the error bars indicate the 99th and 1st percentiles of the distributions. The crosses represent the maximum errors.

**Table S2.** The top five largest deviations between SISSO model predicted zT values and the experimentally measured zT values.

| system | temperature (K) | predicted zT | measured zT | deviation |
|---|---|---|---|---|
| $Ag_{0.55}Cu_{0.45}GaTe_2$ | 730.3 | 1.62 | 2.38 | 0.76 |
| $Ag_{0.55}Cu_{0.45}GaTe_2$ | 769.4 | 1.95 | 2.69 | 0.74 |
| $Ag_{0.6}Cu_{0.4}GaTe_2$ | 769.3 | 1.83 | 2.52 | 0.68 |
| $Ag_{0.6}Cu_{0.4}GaTe_2$ | 730.2 | 1.53 | 2.21 | 0.68 |
| $Ag_{0.525}Cu_{0.475}GaTe_2$ | 730.2 | 1.67 | 2.29 | 0.62 |

**Table S3.** The descriptor components $d_i$, the coefficients $c_i$, and the occurrence number of each component during the 10CV processes for the best SISSO model at the first iteration.

| | descriptor component | coefficient | occurrence number |
|---|---|---|---|
| $d_1$ | $T^3/(AW_B \cdot AW_{C*})$ | 0.93247E-05 | 10 |
| $d_2$ | $(1-C_{A*})/(EN_{C*} \cdot |EN_A-EN_B|)$ | 0.17046E+00 | 1 |
| $d_3$ | $(T/AW_B)^3$ | 0.21914E-03 | 7 |
| $d_4$ | $(T \cdot C_{B*})/(IE_{B*}-IE_C)$ | -0.16632E-02 | 0 |
| $d_5$ | $T \cdot C_{B*} \cdot (AR_{A*}-AR_B)$ | -0.11212E-01 | 0 |
| $d_6$ | $AW_{A*}/(HF_{A*} \cdot HV_B \cdot HV_{B*})$ | 0.82582E-03 | 0 |
| $d_7$ | $C_{B*}^2/HV_{A*}^3$ | -0.15125E+03 | 0 |

**Table S4.** The descriptor components $d_i$, the coefficients $c_i$, and the occurrence number of each component during the 10CV processes for the best SISSO model at the second iteration.

| | descriptor component | coefficient | occurrence number |
|---|---|---|---|
| $d_1$ | $T^3/(AW_B \cdot AW_{C*})$ | 0.28180E-04 | 10 |
| $d_2$ | $T/(AR_B \cdot |EN_A-EN_B|)$ | 0.93246E-04 | 1 |
| $d_3$ | $T^2/(AW_B \cdot AW_{C*})$ | -0.16963E-01 | 3 |
| $d_4$ | $C_{B*}^3 \cdot (AW_{A*}-AW_B)$ | -0.91257E+00 | 6 |
| $d_5$ | $T^2 \cdot (C_{A*}+C_{C*})$ | -0.30352E-05 | 3 |
| $d_6$ | $(HF_A \cdot IE_{B*})/(T \cdot HV_{A*})$ | 0.67027E+02 | 0 |
| $d_7$ | $(AW_{B*}-AW_{C*})/(AW_{A*}-AW_B)$ | -0.15383E-01 | 0 |

**Table S5.** The descriptor components $d_i$, the coefficients $c_i$, and the occurrence number of each component during the 10CV processes for the best SISSO model at the third iteration.

| | descriptor component | coefficient | occurrence number |
|---|---|---|---|
| $d_1$ | $T^2/((1-C_{A*})\cdot AW_B)$ | 0.65004E-04 | 10 |
| $d_2$ | $T\cdot(HV_B-HV_{C*})/(1-C_{A*})$ | 0.12435E-03 | 1 |
| $d_3$ | $(C_{A*}/HV_A)\cdot(HF_A-HF_{A*})$ | 0.12824E+03 | 0 |
| $d_4$ | $(1-C_{A*})^3\cdot AW_B\cdot HV_A$ | 0.15328E-01 | 1 |
| $d_5$ | $(C_{A*}-C_{B*})/(AR_A-AR_C)$ | 0.61174E-01 | 0 |
| $d_6$ | $(AW_{A*}\cdot AW_{C*})/T^2$ | 0.46100E+01 | 0 |
| $d_7$ | $AW_B\cdot C_{B*}\cdot(IE_A-IE_{B*})$ | 0.39042E-02 | 0 |

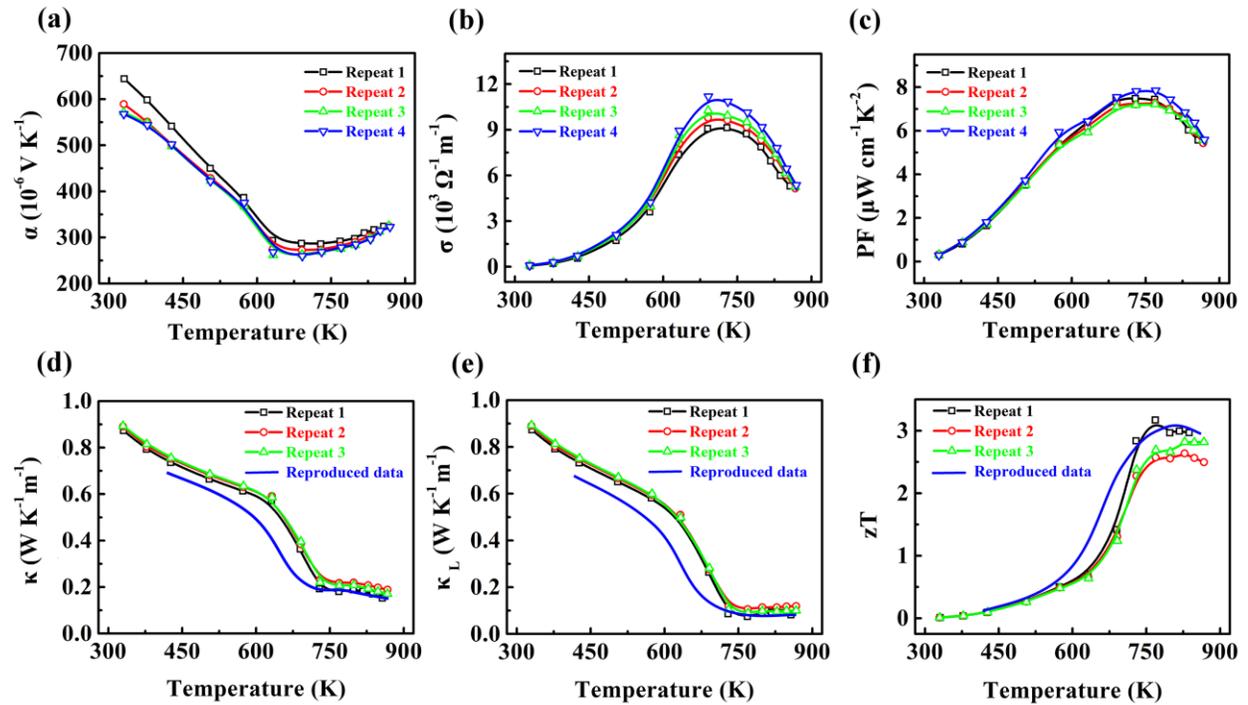

**Figure S3.** The repeated measurements of thermoelectric properties as a function of temperature for $Ag_{0.55}Cu_{0.45}GaTe_2$. (a) Seebeck coefficients ($\alpha$); (b) Electrical conductivities ($\sigma$); (c) Power factors (PF); (d) Total thermal conductivities ($\kappa$); (e) Lattice thermal conductivities ($\kappa_L$); (f) Figure of merit (zT).

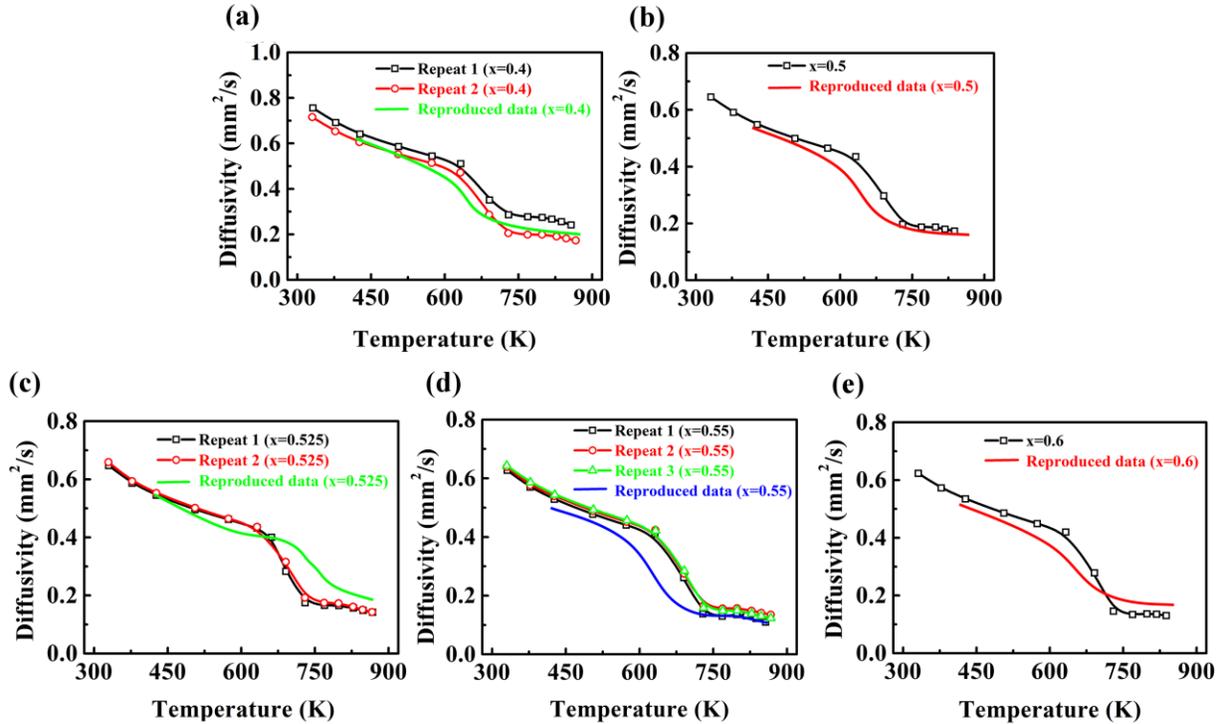

**Figure S4.** Thermal diffusivity as a function of temperature for $Cu_{1-x}Ag_xGaTe_2$ (x=0.4-0.6). (a) x=0.4; (b) x=0.5; (c) x=0.525; (d) x=0.55; (e) x=0.6.

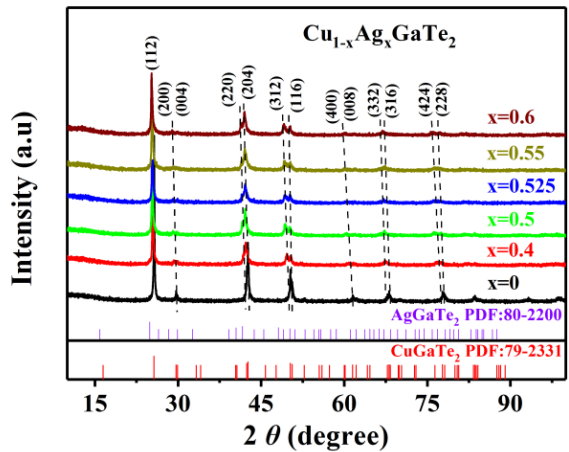

**Figure S5.** Powder XRD patterns for $Cu_{1-x}Ag_xGaTe_2$ (x=0-0.6). The XRD peaks show a shifting trend with increasing x.

**Supplementary References**

**Supplementary Appendix**

Testing report for thermal diffusivity from The Center for Materials Research and Analysis in Wuhan University of Technology.

编号：20210008

# 武汉理工大学材料研究与测试中心

# 数 据 报 告

样品名称：Cu(Ag)GaTe$_2$

送样单位：宁波工程学院（Ningbo University of Technology）

通讯地址：浙江省宁波市江北区风华路 201 号

检测单位（盖章）：武汉理工大学材料研究与测试中心

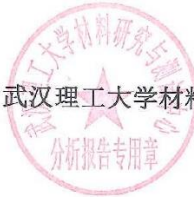

发送日期　2021　年　1　月　11　日

# 检 测 数 据

报告编号：20210008　　　　　　　　共 1 页 第 1 页（其中图 0 页，表 1 页）

| 样品编号 | 202100081701~05 | | 检验类别 | 委托送样 |
|---|---|---|---|---|
| 样品名称 | $Cu(Ag)GaTe_2$ | | | |
| 样品状态 | 黑色固体 | 接样时间 2021/1/5 | 检测时间 | 2021/1/6~1/7 |
| 检测项目 | 热扩散系数 | | | |
| 参考依据 | GB/T 22588－2008 闪光法测量热扩散系数或导热系数 | | | |
| 检测设备 | 激光导热仪(LFA457) | | 设备编号 | 1604780S |
| 检测条件 | 室温:22℃，湿度:32%RH；检测温度范围：室温～580℃，激光电压：1538V，气氛：Ar，流量：80mL/min。 | | | |

检测结果

| 检测项目<br>样品编号（名称） | 热扩散系数（mm²/s） | | | | | | |
|---|---|---|---|---|---|---|---|
| | 室温 | 150℃ | 300℃ | 400℃ | 500℃ | 550℃ | 580℃ |
| 202100081701<br>$(CuGaTe_2)_{0.6}(AgGaTe_2)_{0.4}$ | 0.80 | 0.61 | 0.49 | 0.26 | 0.23 | 0.22 | 0.20 |
| 202100081702<br>$(CuGaTe_2)_{0.5}(AgGaTe_2)_{0.5}$ | 0.68 | 0.53 | 0.42 | 0.21 | 0.18 | 0.17 | 0.16 |
| 202100081703<br>$(CuGaTe_2)_{0.475}(AgGaTe_2)_{0.525}$ | 0.72 | 0.55 | 0.43 | 0.41 | 0.28 | 0.22 | 0.20 |
| 202100081704<br>$(CuGaTe_2)_{0.45}(AgGaTe_2)_{0.55}$ | 0.65 | 0.50 | 0.35 | 0.16 | 0.14 | 0.12 | 0.12 |
| 202100081705<br>$(CuGaTe_2)_{0.4}(AgGaTe_2)_{0.6}$ | 0.66 | 0.50 | 0.39 | 0.22 | 0.19 | 0.17 | 0.16 |

以下空白

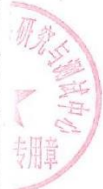

检测人：　　　　校核人：　　　　报告批准人：

签发日期 2021 年 1 月 11 日

声明：本实验数据和结果仅对来样提供参考数据，不作为第三方检测机构的证明或公证使用。
本中心不承担由此带来的任何相关法律责任。

Center for Materials Research and Analysis in Wuhan University of Technology.

# TEST   REPORT

**Sample Name:** Cu(Ag)GaTe$_2$

**Client:** Ningbo University of Technology

**Address:** No. 201, Fenghua Road, Jiangbei District, Ningbo, Zhejiang.

**Inspection Agency:** Center for Materials Research and Analysis in Wuhan University of Technology

# TEST   DATA

No. 20210008          Page 1 of 1

| | | | | | |
|---|---|---|---|---|---|
| Sample No. | 202100081701~05 | | Test Type | Sample delivery | |
| Name | $Cu(Ag)GaTe_2$ | | | | |
| Sample Status | Black solid | Receiving Date | January 5, 2021 | Testing Date | January 6~7, 2021 |
| Test Item | Thermal diffusivity | | | | |
| Reference Standards | GB/T 22588-2008    Laser Flash Method to measure thermal diffusivity or thermal conductivity | | | | |
| Instrument | LFA457 | | Instrument No. | 1604780S | |
| Test Conditions | Room temperature (RT):  22 ℃,  Humidity: 32% RH, Temperature range: RT~580 ℃, Laser voltage: 1538 V, Atmosphere: Ar, Gas flow: 80 mL/min。 | | | | |

**Test Results**

| Test Item  Sample No. | Thermal Diffusivity (mm²/s) | | | | | | |
|---|---|---|---|---|---|---|---|
| | RT | 150℃ | 300℃ | 400℃ | 500℃ | 550℃ | 580℃ |
| 202100081701 $(CuGaTe_2)_{0.6}(AgGaTe_2)_{0.4}$ | 0.80 | 0.61 | 0.49 | 0.26 | 0.23 | 0.22 | 0.20 |
| 202100081702 $(CuGaTe_2)_{0.5}(AgGaTe_2)_{0.5}$ | 0.68 | 0.53 | 0.42 | 0.21 | 0.18 | 0.17 | 0.16 |
| 202100081703 $(CuGaTe_2)_{0.475}(AgGaTe_2)_{0.525}$ | 0.72 | 0.55 | 0.43 | 0.41 | 0.28 | 0.22 | 0.20 |
| 202100081704 $(CuGaTe_2)_{0.45}(AgGaTe_2)_{0.55}$ | 0.65 | 0.50 | 0.35 | 0.16 | 0.14 | 0.12 | 0.12 |
| 202100081705 $(CuGaTe_2)_{0.4}(AgGaTe_2)_{0.6}$ | 0.66 | 0.50 | 0.39 | 0.22 | 0.19 | 0.17 | 0.16 |

**Opeator**: RongHui Zhuo      **Verifier**: XinYa Yang      **Approver**: YanYuan Qi

January 11, 2021

Statement: The experimental data and results only provide reference data for the incoming samples, and are not used as a certification or notarization by a third-party testing agency. The center does not assume any related legal responsibilities resulting from this.